\documentclass[aps,prl,twocolumn,superscriptaddress]{revtex4}
\usepackage{graphicx}
\usepackage{subfigure}
\usepackage{epstopdf}
\usepackage{amsmath}
\usepackage{amssymb}
\usepackage{amsfonts}
\usepackage{mathrsfs}
\usepackage{theorem}
\usepackage{bm}
\usepackage{url}
\usepackage[T1]{fontenc}
\usepackage{csquotes}
\MakeOuterQuote{"}

\usepackage{algorithm}
\usepackage{algorithmicx}
\usepackage{algpseudocode}

\usepackage{dcolumn}
\usepackage{color}

\definecolor{ngreen}{rgb}{0.2,0.6,0.2}

\definecolor{ngold}{rgb}{0.7,0.6,0.2}

\def\vec#1{\mathbf{#1}} %% overiding the original command

\newcommand{\tr}{\operatorname{tr}}

\newcommand{\rmi}{\mathrm{i}}

\newcommand{\be}{\begin{equation}}
\newcommand{\ee}{\end{equation}}
\newcommand{\ba}{\begin{align}}
\newcommand{\ea}{\end{align}}

\def\<{\langle}  %% overiding the original command \<
\def\>{\rangle}  %% overiding the original command \>
\newcommand{\ket}[1]{| #1\>}
\newcommand{\bra}[1]{\< #1|}

\def\outer#1#2{|#1\>\<#2|}       %% overiding the original command \outer

\def\eqref#1{\textup{(\ref{#1})}}  %% overiding the original command \eqref
\newcommand{\eref}[1]{Eq.~\textup{(\ref{#1})}}

\newcommand{\esref}[1]{Eqs.~\textup{(\ref{#1})}}

\newcommand{\fref}[1]{Fig.~\ref{#1}}

\newcommand{\cref}[1]{Conjecture~\ref{#1}}
\newcommand{\Cref}[1]{Conjecture~\ref{#1}}

\begin{document}

\title{Experimental realization of self-guided quantum process tomography}
\author{Zhibo Hou}
\thanks{These authors contributed equally to this work.}
\affiliation{CAS Key Laboratory of Quantum Information, University of Science and Technology of China, Hefei 230026, China}
\affiliation{CAS Center for Excellence in Quantum Information and Quantum Physics, University of Science and Technology of China, Hefei 230026, China}
\author{Jun-Feng Tang}
\thanks{These authors contributed equally to this work.}
\affiliation{CAS Key Laboratory of Quantum Information, University of Science and Technology of China, Hefei 230026, China}
\affiliation{CAS Center for Excellence in Quantum Information and Quantum Physics, University of Science and Technology of China, Hefei 230026, China}
\author{Christopher Ferrie}
\email{christopher.ferrie@uts.edu.au}
\affiliation{University of Technology Sydney, Centre for Quantum Software and Information, Ultimo NSW 2007, Australia}
\author{Guo-Yong Xiang}
\email{gyxiang@ustc.edu.cn}
\affiliation{CAS Key Laboratory of Quantum Information, University of Science and Technology of China, Hefei 230026, China}
\affiliation{CAS Center for Excellence in Quantum Information and Quantum Physics, University of Science and Technology of China, Hefei 230026, China}
\author{Chuan-Feng Li}
\affiliation{CAS Key Laboratory of Quantum Information, University of Science and Technology of China, Hefei 230026, China}
\affiliation{CAS Center for Excellence in Quantum Information and Quantum Physics, University of Science and Technology of China, Hefei 230026, China}
\author{Guang-Can Guo}
\affiliation{CAS Key Laboratory of Quantum Information, University of Science and Technology of China, Hefei 230026, China}
\affiliation{CAS Center for Excellence in Quantum Information and Quantum Physics, University of Science and Technology of China, Hefei 230026, China}

\date{\today}

\begin{abstract}
Characterization of quantum processes is a preliminary step necessary in the development of quantum technology. The conventional method uses standard quantum process tomography, which requires $d^2$ input states and $d^4$ quantum measurements for a $d$-dimensional Hilbert space. These experimental requirements are compounded by the complexity of processing the collected data, which can take several orders of magnitude longer than the experiment itself. In this paper we propose an alternative self-guided algorithm for quantum process tomography, tuned for the task of finding an unknown unitary process. Our algorithm is a fully automated and adaptive process characterization technique. The advantages of our algorithm are: inherent robustness to both statistical and technical noise; requires less space and time since there is no post-processing of the data; requires only a single input state and measurement; and, provides on-the-fly diagnostic information while the experiment is running.  
Numerical results show our algorithm achieves the same $1/n$ scaling as standard quantum process tomography when $n$ uses of the unknown process are used. We also present experimental results wherein the algorithm, and its advantages, are realized for the task of finding an element of $SU(2)$. 
\end{abstract}

\date{\today}
\maketitle

\emph{Introduction}---The advantages of quantum technologies grow rapidly with the size of the quantum systems involved. Naturally, then, it is of interest to produce and verify larger and larger quantum states, with more resourceful states being generally more difficult to both realize and validate. For example, genuine multipartite entangled states have been prepared with 12 superconducting qubits in linear cluster states \cite{gong2019genuine}, 14 trapped ions \cite{MonzSBC11} and 18 photonic qubits \cite{wang201818} in Greenberger-Horne-Zeilinger (GHZ) states. However, our capability to characterizing large quantum systems themselves and their dynamics are left far behind \cite{Haff05scalable,Silv11practical,Smol12efficient,hou2016full,shang2017superfast}. The conventional method,  quantum tomography, performs an informationally complete positive valued measurement (IC-POVM) and outputs an estimate based on all outcomes \cite{Pari04quantum,Haya05asymptotic}. However this process demands exponential cost of both quantum measurements, data storage and computational time  with the size of the quantum systems \cite{hou2016full}.

Many methods have been proposed to solve this problem with varying degrees of success. For example, compressed-sensing-like estimates can realize quadratic improvement when the rank of the state is constant \cite{GrosLFB10,Weit12experimental,Kale15quantum}, while permutation-invariant tomography achieves greater efficiency at the cost of stronger assumptions \cite{Toth10permutationally}. While it is impossible to eliminate the exponential scaling of quantum tomography without strong assumptions, for a given system size we can realize large practical gains in efficiency with methods that seek to optimize the figures of merit we are interested in. Recently, a self-guided learning algorithm was proposed to learn quantum states by adaptive measurements without the need of post processing at all \cite{ferrie2014self}. This method provides many practical advantages that have not been considered in previous tomographic protocols. For example, it does not require storing large data sets, it does not require post processing of data, and it is robust to technical noise and random calibration errors. These advantages have been experimentally demonstrated in characterizing photonic states in optical systems \cite{chapman2016experimental}.

Characterizing the dynamics experienced by quantum states is more demanding than estimating the quantum states alone. For static processes, this is typically called quantum process tomography (QPT) \cite{chuang1997prescription,Poyatos97complete,d2003quantum,Pari04quantum,Mohseni08quantum,Pogorelov17experimental}. The characterization of a quantum processes requires $O(d^4)$ parameters for a $d-$dimensional system, as opposed to the state estimation problem which requires $O(d^2)$. Many quantum state estimation protocols have been straightforwardly generalized to quantum processes. However, techniques that rely directly on metrics, including adaptive protocols, fail because the natural metrics on processes are completely different than those of states. This includes the self-guided quantum state tomography protocol of \cite{ferrie2014self}.

Given its many advantages, then, it is natural to extend the self-guided quantum tomography algorithm for this more difficult problem of characterizing quantum processes. In this paper we develop such a self-guided quantum process tomography (SGQPT) algorithm, and experimentally demonstrate it to find a unitary $SU(2)$ processes. Our algorithm only needs one input state, which is maximally entangled with an ancilla, and one Bell measurement performed on the system qubit and the ancilla. This reduces the complexity of state preparation and quantum measurements significantly, compared with the spanning set of input states and measurements required for conventional quantum tomography. Simulations to extract asymptotic behavior demonstration that our estimation algorithm achieves the same precision as standard quantum process tomography, which is expected. Importantly, this comes at the great benefit of reduced space and computational complexity and well has inherent robustness to noise. The algorithm can be applied generally to any metric, though here we only specify it for the fidelity metric on $SU(d)$. As the data obtained is informationally complete, statistical techniques can be added to self-guided protocols to extract additional information \cite{Granade_2017}.

\emph{Self-guided quantum process tomography}---To characterize a quantum process, standard quantum process tomography proceeds as follows: send a known spanning set of input states through the unknown channel and perform quantum state tomography on the output. From these input-output relations, reconstruct a valid quantum process. When the process is assumed to be unitary, a quadratic improvement in complexity can be realized, requiring $O(d^2)$ different measurements for each of $O(d)$ input pure states \cite{PhysRevA.90.012110}. After all the data is recorded, the resulting linear algebra problem requires at best $O(d^6)$ complexity arising from inverting a $d^2\times d^2$ matrix. Here we propose and demonstrate SGQPT, an algorithm that uses a single input state and measurement and requires only the $O(d^2)$ numbers to store and add together running estimates of the unknown unitary.

SGQPT uses a similar philosophy as self-guided quantum \emph{state} tomography toward quantum learning by treating the estimation problem as a directed search optimization problem. Suppose $U$ is the $SU(d)$ operator to be estimated and $V$ is another element of $SU(d)$ which can be controlled. When an input state $\ket{\psi}$ is evolved by $U$, followed by $V^\dagger$, we have an output state $V^\dagger U\ket{\psi}$. Clearly, since we control $V$, we have succeeded in learning $U$ when $V=U$, in which case $V^\dagger U\ket{\psi}=\ket{\psi}$. A naive idea is to measure on the input state $\outer{\psi}{\psi}$ and if we get probability one, then we say we have learned $V=U$. However, this is not true because $\bra{\psi}V^\dagger U\ket{\psi}=1$ can be satisfied by both $I$ and $V^\dagger U=2\outer{\psi}{\psi}-I$. 

This problem can be solved when assisted by a maximally entangled ancilla, i.e. $\ket{\psi}=\frac{1}{\sqrt{d}}\sum_{j=1}^{d}\ket{jj}$. Allowing the unknown unitary and $V^\dagger$ to act on half of this entangled state and measuring against the same state results in an outcome probability,
\begin{equation}\label{eq:f probabiltiy}
f(V)=\bra{\psi}V^\dagger U\otimes I\ket{\psi}=\frac{1}{d}\tr{V^\dagger U}.
\end{equation}
In this case, getting measurement outcome $\outer{\psi}{\psi}$ with probability 1 requires $V^\dagger U=I$. Accordingly, we can treat the task of learning $U$ as one of optimizing $f$. In other words, we modify $V$ so that the measurement probability of $\outer{\psi}{\psi}$ reaches 1 as fast as possible.

Since we cannot measure probabilities (in our finite lifespans anyway), we look toward \emph{stochastic approximation} optimization techniques. In particular, we use the simultaneous perturbation stochastic approximation (SPSA) algorithm to update $V$ \cite{spall1992multivariate,ferrie2014self}. SPSA is an iterative optimization technique which uses only two (noisy) function calls per iteration to estimate the gradient. To be specific, in step $k$ we first generate a random direction $\Delta_k$ to search; then the gradient $G_k$ is calculated in that direction via 
\begin{equation}\label{eq:gradient}
G_{k}=\frac{\hat f\left(V_{k}+\delta_{k} \Delta_{k}\right)-\hat f\left(V_{k}-\delta_{k} \Delta_{k}\right)}{2 \delta_{k}} \Delta_{k},
\end{equation}
where $\hat f$ is an estimate of $f$ drawn from experimental samples, which can be subjected to quite general sources of statistical and technical noise.
The learned operation is updated by
\begin{equation}\label{eq:update V}
V_{k+1}=V_k+g_kG_k.
\end{equation}
The parameters $\delta_k$ and $g_k$ control the convergence speed and we take the following empirical formula
\begin{equation}\label{eq:iteration}
\begin{aligned}
\delta_k = {\delta_0 \over (k+1)^\gamma},\ \ g_k = {g_0 \over (k+1+A)^\alpha}.
\end{aligned}
\end{equation}
Three of the parameters are chosen as fixed values in the simulations $\delta_0=0.2,g_0=2$ and $A=0$ while the other two, $\gamma$ and $\alpha$ depend on the problem studied. 
For the random direction $\Delta_k$, each of its elements chooses either 1 or $-1$ with the same probability 0.5.  As this direction is chosen randomly, it guarantees an unbiased estimation of the gradient and avoids all the measurements necessary to estimation the exact gradient.

In our algorithm, we only need to store one current estimate $V_k$, one random direction $\Delta_k$, and two measurement results. After several iterations of learning, $V_k$ is directly our estimate of the unknown operator $U$. 

\begin{figure}[t]
	\center{\includegraphics[scale=0.5]{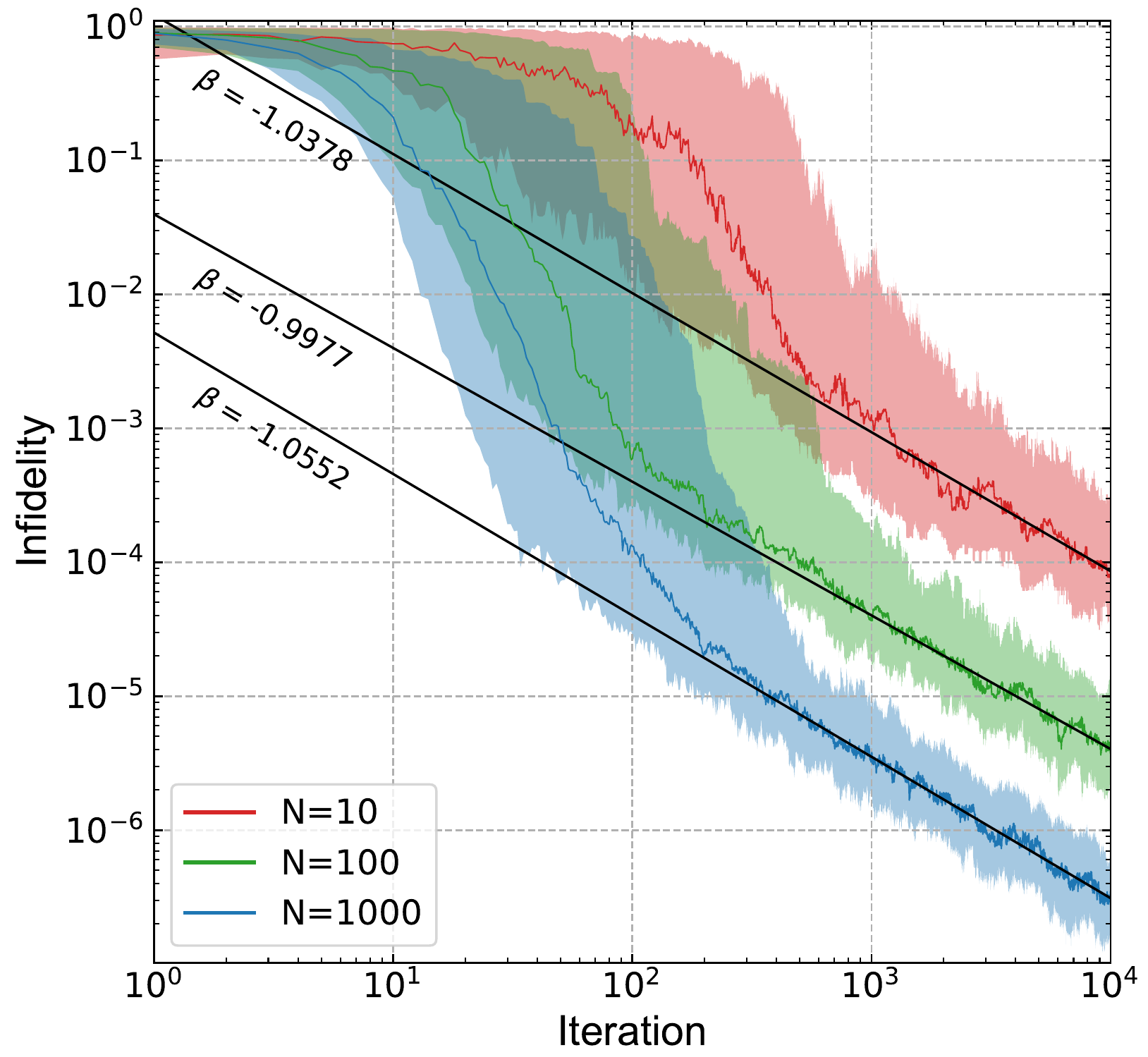}}
	\caption{\label{fig:numerical no noise}
		Numerical results without noise. The solid lines are the median performance of SGQPT over 100 randomly (according to Haar measure) chosen $SU(2)$ operators. We use up to $10^5$ iterations in the simulation and the scaling behavior of infidelity is fitted with respect to the number of iterations. The parameters in the algorithm are optimized to achieve the best fitting scaling behavior, specified in the figure. The optimized parameters are $\alpha=0.92, \gamma=0.42$. In each iteration, two controls are added and both are measured $N$ times.
	}
\end{figure}
\begin{figure}[h]
	\center{\includegraphics[scale=0.5]{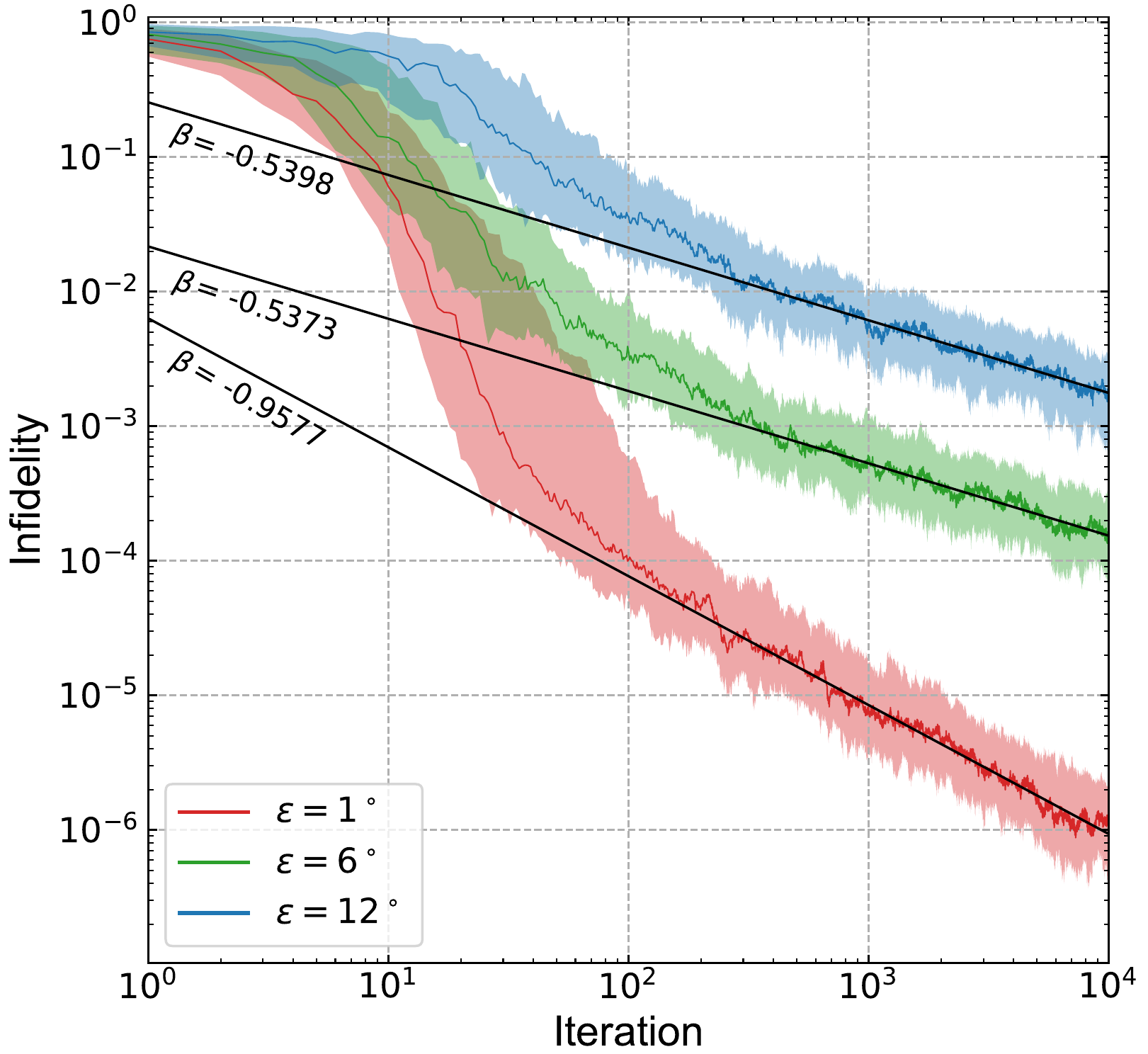}}
	\caption{\label{fig:numerical noise}
		Numerical results with three levels of random noise. Noise is added to the rotation angles of the three wave plates in the implementation of the learning control $V$ by choosing a random value in the uniform interval $(-\epsilon,\epsilon)$, where $\epsilon$ is specified in the legend. Parameters in our algorithm are numerically optimized to achieve the best scaling behavior of the median performance with respect to the number of iterations. The optimized parameters are $\alpha=0.94, \gamma=0.21$. In each iteration, the measurements are repeated 1000 times for both controls.
	}
\end{figure}
\begin{figure*}[t]
	\center{\includegraphics[scale=1.2]{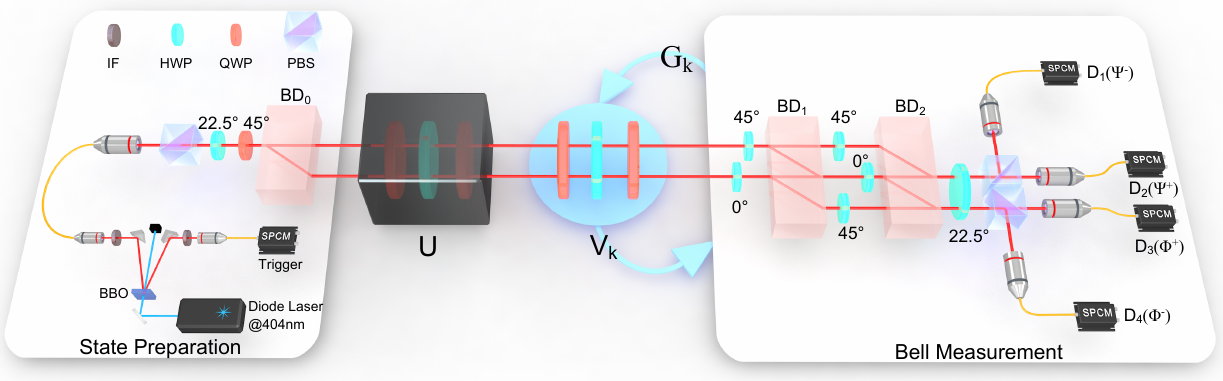}}
	\caption{\label{fig:setup}
		Experimental setup. There are four modules in the experimental setup: Bell state preparation, unknown evolution, learning control evolution and Bell measurement. The system and ancilla qubits are encoded on the polarization and path degree of freedom, respectively. The learning controls in each iteration are automatically realized by a Python program. Bell measurements on system and ancilla qubits are implemented by a three-step quantum walk.
	}
\end{figure*}
\emph{Numerical results}---In numerical simulations, we seek to estimate an unknown element of $SU(2)$. We take the following parameterization
\begin{equation}\label{eq:parameterization}
U=e^{\rmi\alpha\vec{n}(\theta,\phi)\cdot\vec{\sigma}}
\end{equation}
with $\vec{n}(\theta,\phi)=(\sin\theta\cos\phi,\sin\theta\sin\phi,\cos\theta)$ and $\vec{\sigma}=(\sigma_1,\sigma_2,\sigma_3)$ are the Pauli operators. Any $SU(2)$ operator is uniquely determined by the set of three parameters, i.e., $(\alpha,\theta,\phi)$. This vector of three parameters in turn can be used to represent the operators $V_k$, directions $\Delta_k$ and also gradients $G_k$ in \esref{eq:gradient} and (\ref{eq:update V}). When the matrix form is needed as in \eref{eq:f probabiltiy}, it can be obtained from its vector form according to \eref{eq:parameterization}.

We first consider the performance of our self-learning algorithm assuming ideal experiments without experimental errors. In the numerical simulation, the only noise is shot noise in the estimation of the probability $f(V_k\pm \beta_k\Delta_k)$ due to the finite number of copies $N$. Two parameters in the SGQPT algorithm are chosen as $\gamma=0.42$ and $\alpha=0.92$ after numerical optimization. Starting from initial values $(\alpha_0,\theta_0,\phi_0)=(\pi/4,\pi/2,\pi)$, the learning process is iterated up to $10^5$ times and the copies $N$ to estimate the two probabilities range from $N=10$ to $1000$ in each iteration. To see an overall performance we selected 100 $SU(2)$ operators, randomly chosen according to the Haar measure.   The figure of merit is chosen as infidelity, which is defined as $1-f(V_k)$. The numerical results are presented in \fref{fig:numerical no noise}. As the performance depends on the specific operator, typical performances are shown by shading area, which covers the 25\% to 75\% range from the best to the worst for the 100 operators. For all three cases of $N$, SGQPT converges very well to the true state. The scaling of SGQPT in all cases is inferred from a fit of median performance to $ck^\beta$, and $\beta\approx -1$ with respect to the number of iterations, achieving the same scaling behavior as expected from standard quantum process tomography, when the latter is performed optimally.

We have also numerically simulated the performance of SGQPT in the scenario with experimental errors. Since experiments are never perfect, a good measurement scheme should be robust against unforeseen experimental errors. Here we simulate experimental errors  in a photonic optical setup, where the system qubit is physically encoded on the polarization degree of freedom. For this kind of physical encoding, controls are implemented experimentally by rotating two quarter-wave plates (QWPs) and one half-wave plate (HWP) \cite{simon1990minimal,simon2012hamilton}. We modeled random errors on the rotation angles of these three wave plates due to limited precision of the rotation stage. Noise is added to each wave plate by choosing a random value in the uniform interval $(-\epsilon,\epsilon)$ and rotating the wave plate by that number of degrees, unknown to the algorithm of course. To highlight the effects of technical errors, we choose $N=1000$ such that shot noise is present but not dominant. Numerical results in \fref{fig:numerical noise} show that the effects of simulated experimental noise on the scaling behavior of SGQPT are not severe. Indeed, for small experimental errors, there is only a slight degradation in performance.  Although the scaling behavior is suboptimal in the high-noise regime, further algorithmic parameter optimization in this limit was not the focus and could lead to better performance. We conclude that SGQPT demonstrates good overall robustness to experimental errors in simulation.

\emph{Experimental setup}---Our SGQPT experiment has four modules: Bell state preparation, unknown evolution, learning control and Bell measurement, as shown in \fref{fig:setup}. In the preparation module, we use a 40-mW V-polarized beam at 404~nm to pump a 1-mm-long BBO crystal, cut for type-\uppercase\expandafter{\romannumeral1} phase-matched spontaneous parametric down-conversion (SPDC) process,  to generate a pair of photons at 808 nm in H-polarization. One photon is measured after transmitting through an interference (IF) to herald the other photon as a single photon source \cite{Kwia99ultrabright}. Then a combination of polarizing beam splitter (PBS), HWP and QWP prepares the photon to the state ${1 \over \sqrt{2}}(\ket{H}+\ket{V})$. After passing through the beam displacer, the system qubit in the polarization degree of the photon with basis $\ket{H}$ and $\ket{V}$ is maximally entangled with the path degree of the photon with basis $\ket{\rm up}$ and $\ket{\rm down}$, i.e., ${1 \over \sqrt{2}}(\ket{H,{\rm up}} + \ket{V,{\rm down}})$.  Then the polarization qubit undergoes the unknown evolution $U$, which consists of two QWPs and one HWP. The learning control module can realize arbitrary $SU(2)$ operator $V_k$ by rotating the rotation angles of two QWPs and one HWP \cite{simon1990minimal,simon2012hamilton}, which are installed on electrically motorized stages for automatic learning control. Finally in the Bell measurement module, the states are measured by complete Bell measurements, which outputs the result of the probability function in \eref{eq:f probabiltiy}.

\begin{figure}[t]
	\center{\includegraphics[scale=0.5]{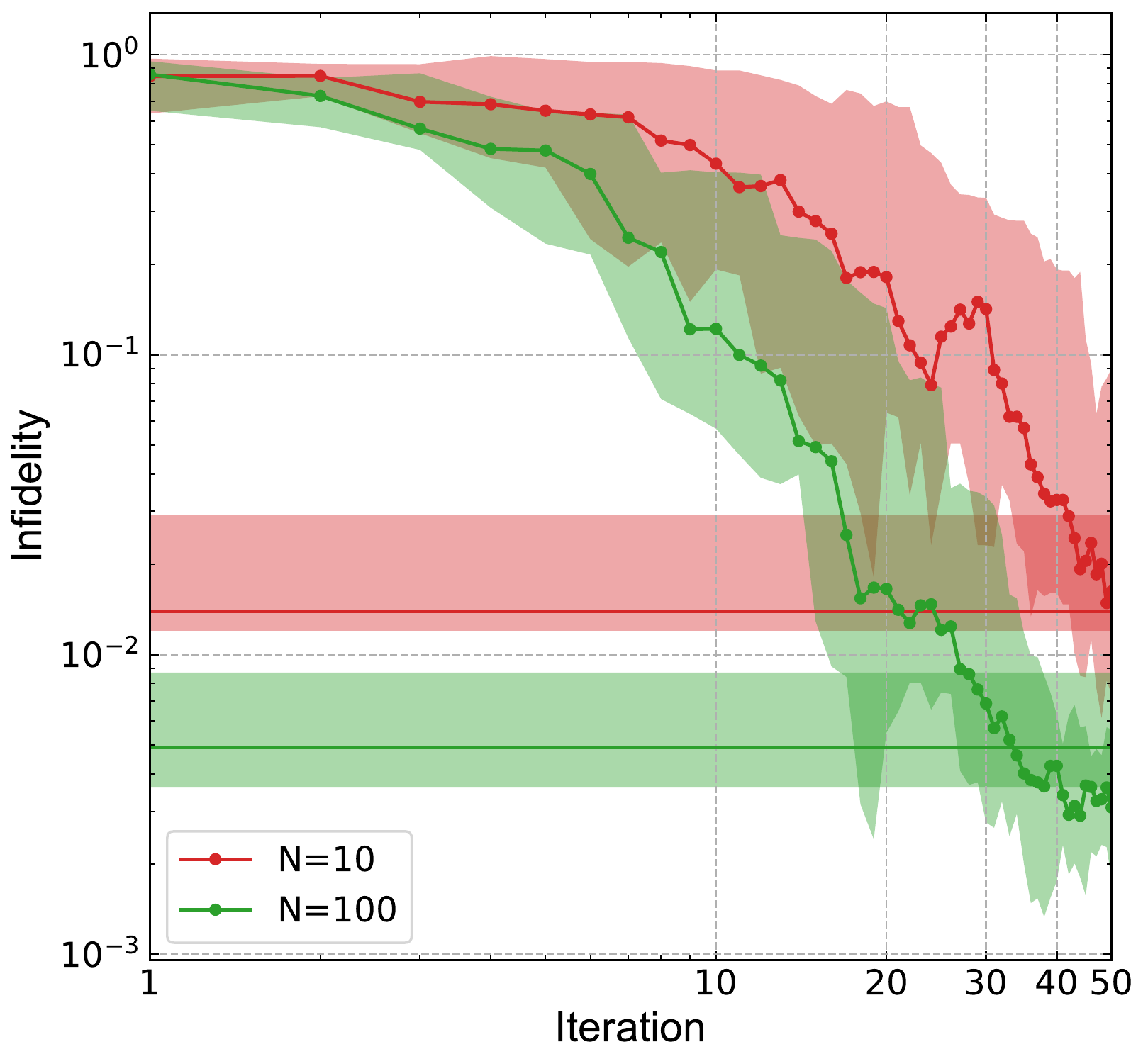}}
	\caption{\label{fig:exp no noise}
	Experimental results without noise. We only perform 50 iterations of learning and $N=10, 100$ repeated measurements for each iteration due to limited time. Parameters $\alpha=0.85$ and $\gamma=0.06$ are optimized to achieve the best median performance over 20 randomly chosen $SU(2)$ operators after 50 iterations of learning. The horizontal lines are simulated median infidelities of QPT with the same total number of photons used in SGQPT with 50 iterations, respectively. The simulation results of quantum process tomography are performed with 6 eigenstates of 3 Pauli matrices as probe states and performing
	3 Pauli measurements for each probe state.
	}
\end{figure}

\emph{Experimental results}---In the experiment, we first considered the scenario without experimental errors. In total we implemented 50 iterations of learning and the copies for each measurement in each iteration are chosen as $N=10$ (red in \fref{fig:exp no noise}) and 100 (green in \fref{fig:exp no noise}), respectively. The algorithm parameters used in the SGQPT experiment  were $\alpha=0.85,\gamma=0.06$. The experimental results for the performance of out SGQPT are plotted in \fref{fig:exp no noise}. For comparison, we also plotted the performance (straight solid line) of standard quantum process tomography with the largest number of copies for iterations of 50 in our case. For example, at $N=100$, the median infidelity achieved by SGQPT after 50 iterations of learning is $3.1\times10^{-3}$, slightly superior to the median infidelity $4.9\times10^{-3}$ of QPT with the same number of copies. It shows that the median performance of our SGQPT can achieve the same performance of QPT in ideal experiments.

\begin{figure}[t]
\center{\includegraphics[scale=0.5]{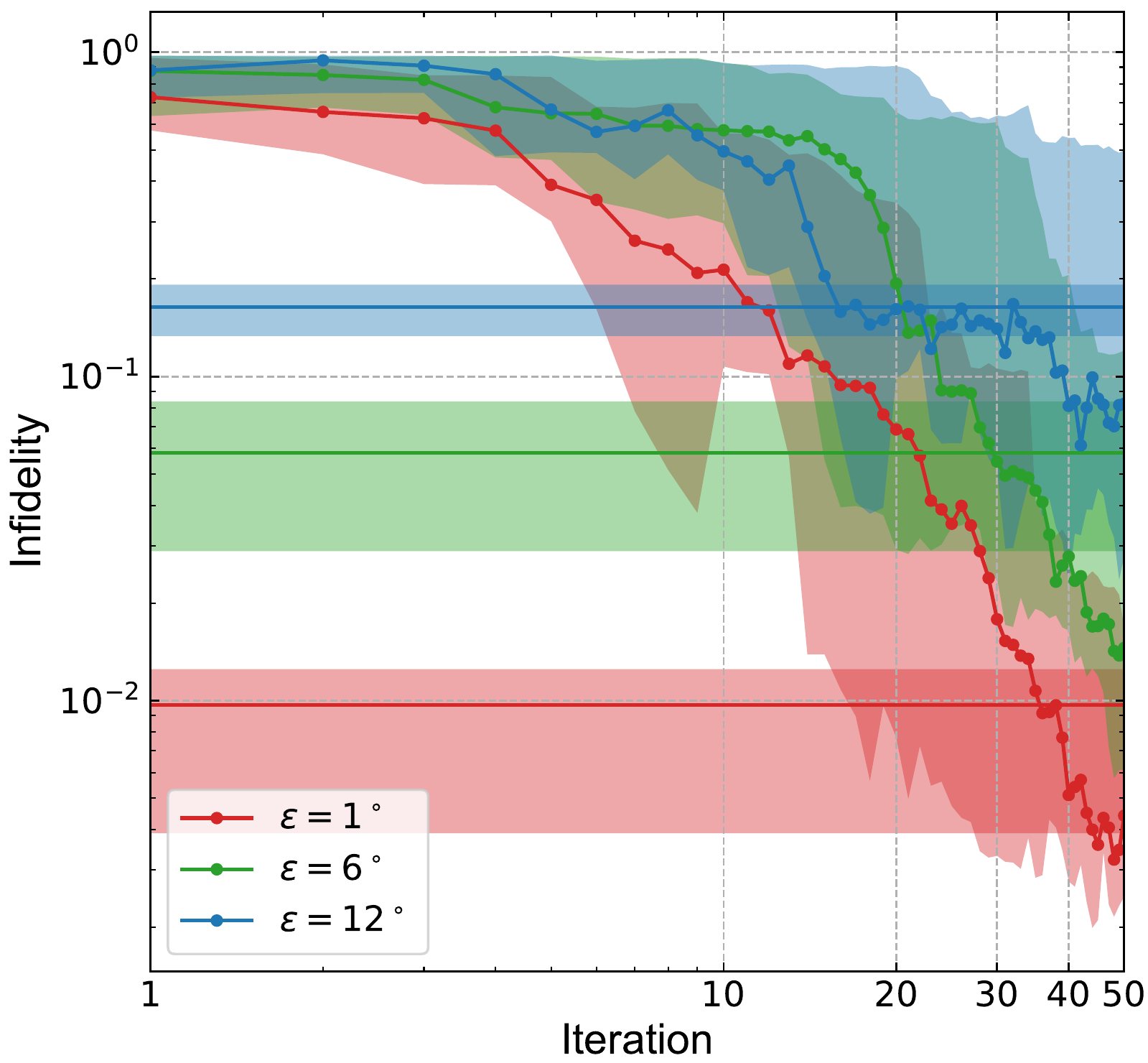}}
\caption{\label{fig:noisecompare}
Experimental results under three levels of random experimental errors. The lines with dots show the median infidelities of 20
randomly chosen $SU(2)$ unitary processes. The shaded regions represent the interquartile range of infidelities. The horizontal lines 	are simulated data, which represent median infidelities of QPT with same levels of error for the same 20 processes above. The noise level means 	the uncertainty of rotation angle of each electric motor stage in degree. The total number of photons used in each tomography for each chosen unitary process is $10^4$.
}
\end{figure}

We also implemented our SGQPT algorithm in the presence of engineered experimental errors. Again, noise is added to the three wave plates in the implementation of the learning control $V$ by choosing a random value in the uniform interval $(-\epsilon,\epsilon)$. The parameters in SGQPT algorithm are chosen the same values as those in the ideal experiments and repeated number of measurements is $N=100$. We chose three levels of random noise on the rotation angles of the wave plates, which are $\epsilon = 1,6,12$ degrees. The experimental results, shown in \fref{fig:noisecompare}, demonstrate the expected convergent behavior in the presence of noise and the median infidelity with 50 iterations of learning is degraded from $3.1\times10^{-3}$ in above ideal experiments to $4.4\times10^{-3}$, $1.45\times10^{-2}$ and $8.23\times10^{-2}$ for the three levels of noise, respectively. For comparison, we also simulated the performance of standard quantum process tomography with copies equal to that used in SGQPT at iteration 50 and the median infidelity is degraded from $4.9\times10^{-3}$ to $9.7\times10^{-3}$, $5.82\times10^{-2}$ and $1.638\times10^{-1}$.  It shows that advantages of SGQPT over QPT increase with the level of noises, and is thus more robust against noise. 

\emph{Summary}---We extended self-guided quantum tomography from learning quantum states to learning quantum processes. In the measurement step, our method uses a fixed input state and a fixed measurement, significantly reducing the complexity of preparation and measurement. In the data processing step, our algorithm does not require post-processing, which thus not only saves storage space but also is computationally efficient compared with standard quantum process tomography. Numerical and experimental results demonstrate that our algorithm has the same precision as QPT in ideal experiments and is very robust against the technical noise expected in experiments. We expect self-guided quantum process tomography to have promising applications in characterizing the dynamics of larger quantum systems in the future, especially when lack of fully automated and embedded characterization and calibration techniques becomes a bottleneck in the development of quantum technology.

The work at USTC is supported by the National Natural Science Foundation of China under Grants (Nos. 11574291 and 11774334),  the National Key Research and Development Program of China (Grant No.2018YFA0306400 and No.2017YFA0304100), Key Research Program of Frontier Sciences, CAS (No.QYZDY-SSW-SLH003), Anhui Initiative in Quantum Information Technologies and  China Postdoctoral Science Foundation (Grant Nos.2016M602012 and 2018T110618). CF acknowledges funding from the Australian Research Council Discovery Early Career Researcher Award (No. DE160100821).

\bibliographystyle{naturemag}
\bibliography{all_references}

\begin{thebibliography}{10}
\expandafter\ifx\csname url\endcsname\relax
  \def\url#1{\texttt{#1}}\fi
\expandafter\ifx\csname urlprefix\endcsname\relax\def\urlprefix{URL }\fi
\providecommand{\bibinfo}[2]{#2}
\providecommand{\eprint}[2][]{\url{#2}}

\bibitem{gong2019genuine}
\bibinfo{author}{Gong, M.} \emph{et~al.}
\newblock \bibinfo{title}{Genuine 12-qubit entanglement on a superconducting
  quantum processor}.
\newblock \emph{\bibinfo{journal}{Phys. Rev. Lett.}}
  \textbf{\bibinfo{volume}{122}}, \bibinfo{pages}{110501}
  (\bibinfo{year}{2019}).
\newblock
  \urlprefix\url{https://link.aps.org/doi/10.1103/PhysRevLett.122.110501}.

\bibitem{MonzSBC11}
\bibinfo{author}{Monz, T.} \emph{et~al.}
\newblock \bibinfo{title}{14-qubit entanglement: Creation and coherence}.
\newblock \emph{\bibinfo{journal}{Phys. Rev. Lett.}}
  \textbf{\bibinfo{volume}{106}}, \bibinfo{pages}{130506}
  (\bibinfo{year}{2011}).
\newblock
  \urlprefix\url{https://link.aps.org/doi/10.1103/PhysRevLett.106.130506}.

\bibitem{wang201818}
\bibinfo{author}{Wang, X.-L.} \emph{et~al.}
\newblock \bibinfo{title}{18-qubit entanglement with six photons' three degrees
  of freedom}.
\newblock \emph{\bibinfo{journal}{Phys. Rev. Lett.}}
  \textbf{\bibinfo{volume}{120}}, \bibinfo{pages}{260502}
  (\bibinfo{year}{2018}).
\newblock
  \urlprefix\url{https://link.aps.org/doi/10.1103/PhysRevLett.120.260502}.

\bibitem{Haff05scalable}
\bibinfo{author}{H{\"a}ffner, H.} \emph{et~al.}
\newblock \bibinfo{title}{Scalable multiparticle entanglement of trapped ions}.
\newblock \emph{\bibinfo{journal}{Nature}} \textbf{\bibinfo{volume}{438}},
  \bibinfo{pages}{643} (\bibinfo{year}{2005}).

\bibitem{Silv11practical}
\bibinfo{author}{da~Silva, M.~P.}, \bibinfo{author}{Landon-Cardinal, O.} \&
  \bibinfo{author}{Poulin, D.}
\newblock \bibinfo{title}{Practical characterization of quantum devices without
  tomography}.
\newblock \emph{\bibinfo{journal}{Phys. Rev. Lett.}}
  \textbf{\bibinfo{volume}{107}}, \bibinfo{pages}{210404}
  (\bibinfo{year}{2011}).
\newblock
  \urlprefix\url{http://link.aps.org/doi/10.1103/PhysRevLett.107.210404}.

\bibitem{Smol12efficient}
\bibinfo{author}{Smolin, J.~A.}, \bibinfo{author}{Gambetta, J.~M.} \&
  \bibinfo{author}{Smith, G.}
\newblock \bibinfo{title}{Efficient method for computing the maximum-likelihood
  quantum state from measurements with additive gaussian noise}.
\newblock \emph{\bibinfo{journal}{Phys. Rev. Lett.}}
  \textbf{\bibinfo{volume}{108}}, \bibinfo{pages}{070502}
  (\bibinfo{year}{2012}).
\newblock
  \urlprefix\url{https://link.aps.org/doi/10.1103/PhysRevLett.108.070502}.

\bibitem{hou2016full}
\bibinfo{author}{Hou, Z.} \emph{et~al.}
\newblock \bibinfo{title}{Full reconstruction of a 14-qubit state within four
  hours}.
\newblock \emph{\bibinfo{journal}{New Journal of Physics}}
  \textbf{\bibinfo{volume}{18}}, \bibinfo{pages}{083036}
  (\bibinfo{year}{2016}).
\newblock
  \urlprefix\url{https://doi.org/10.1088%2F1367-2630%2F18%2F8%2F083036}.

\bibitem{shang2017superfast}
\bibinfo{author}{Shang, J.}, \bibinfo{author}{Zhang, Z.} \&
  \bibinfo{author}{Ng, H.~K.}
\newblock \bibinfo{title}{Superfast maximum-likelihood reconstruction for
  quantum tomography}.
\newblock \emph{\bibinfo{journal}{Phys. Rev. A}} \textbf{\bibinfo{volume}{95}},
  \bibinfo{pages}{062336} (\bibinfo{year}{2017}).
\newblock \urlprefix\url{https://link.aps.org/doi/10.1103/PhysRevA.95.062336}.

\bibitem{Pari04quantum}
\bibinfo{editor}{Paris, M. G.~A.} \& \bibinfo{editor}{{\v{R}}eh\'{a}\v{c}ek,
  J.} (eds.) \emph{\bibinfo{title}{Quantum State Estimation}}, vol.
  \bibinfo{volume}{649} of \emph{\bibinfo{series}{Lecture Notes in Physics}}
  (\bibinfo{publisher}{Springer}, \bibinfo{address}{Berlin},
  \bibinfo{year}{2004}).

\bibitem{Haya05asymptotic}
\bibinfo{author}{Hayashi, M.}
\newblock \emph{\bibinfo{title}{Asymptotic Theory of Quantum Statistical
  Inference}} (\bibinfo{publisher}{WORLD SCIENTIFIC}, \bibinfo{year}{2005}).
\newblock \urlprefix\url{https://www.worldscientific.com/doi/abs/10.1142/5630}.
\newblock \eprint{https://www.worldscientific.com/doi/pdf/10.1142/5630}.

\bibitem{GrosLFB10}
\bibinfo{author}{Gross, D.}, \bibinfo{author}{Liu, Y.-K.},
  \bibinfo{author}{Flammia, S.~T.}, \bibinfo{author}{Becker, S.} \&
  \bibinfo{author}{Eisert, J.}
\newblock \bibinfo{title}{Quantum state tomography via compressed sensing}.
\newblock \emph{\bibinfo{journal}{Phys. Rev. Lett.}}
  \textbf{\bibinfo{volume}{105}}, \bibinfo{pages}{150401}
  (\bibinfo{year}{2010}).
\newblock
  \urlprefix\url{https://link.aps.org/doi/10.1103/PhysRevLett.105.150401}.

\bibitem{Weit12experimental}
\bibinfo{author}{Liu, W.-T.}, \bibinfo{author}{Zhang, T.},
  \bibinfo{author}{Liu, J.-Y.}, \bibinfo{author}{Chen, P.-X.} \&
  \bibinfo{author}{Yuan, J.-M.}
\newblock \bibinfo{title}{Experimental quantum state tomography via compressed
  sampling}.
\newblock \emph{\bibinfo{journal}{Phys. Rev. Lett.}}
  \textbf{\bibinfo{volume}{108}}, \bibinfo{pages}{170403}
  (\bibinfo{year}{2012}).
\newblock
  \urlprefix\url{https://link.aps.org/doi/10.1103/PhysRevLett.108.170403}.

\bibitem{Kale15quantum}
\bibinfo{author}{Kalev, A.}, \bibinfo{author}{Kosut, R.~L.} \&
  \bibinfo{author}{Deutsch, I.~H.}
\newblock \bibinfo{title}{Quantum tomography protocols with positivity are
  compressed sensing protocols}.
\newblock \emph{\bibinfo{journal}{npj Quantum Information}}
  \textbf{\bibinfo{volume}{1}}, \bibinfo{pages}{15018} (\bibinfo{year}{2015}).
\newblock \urlprefix\url{http://dx.doi.org/10.1038/npjqi.2015.18}.

\bibitem{Toth10permutationally}
\bibinfo{author}{T\'oth, G.} \emph{et~al.}
\newblock \bibinfo{title}{Permutationally invariant quantum tomography}.
\newblock \emph{\bibinfo{journal}{Phys. Rev. Lett.}}
  \textbf{\bibinfo{volume}{105}}, \bibinfo{pages}{250403}
  (\bibinfo{year}{2010}).
\newblock
  \urlprefix\url{https://link.aps.org/doi/10.1103/PhysRevLett.105.250403}.

\bibitem{ferrie2014self}
\bibinfo{author}{Ferrie, C.}
\newblock \bibinfo{title}{Self-guided quantum tomography}.
\newblock \emph{\bibinfo{journal}{Phys. Rev. Lett.}}
  \textbf{\bibinfo{volume}{113}}, \bibinfo{pages}{190404}
  (\bibinfo{year}{2014}).
\newblock
  \urlprefix\url{https://link.aps.org/doi/10.1103/PhysRevLett.113.190404}.

\bibitem{chapman2016experimental}
\bibinfo{author}{Chapman, R.~J.}, \bibinfo{author}{Ferrie, C.} \&
  \bibinfo{author}{Peruzzo, A.}
\newblock \bibinfo{title}{Experimental demonstration of self-guided quantum
  tomography}.
\newblock \emph{\bibinfo{journal}{Phys. Rev. Lett.}}
  \textbf{\bibinfo{volume}{117}}, \bibinfo{pages}{040402}
  (\bibinfo{year}{2016}).
\newblock
  \urlprefix\url{https://link.aps.org/doi/10.1103/PhysRevLett.117.040402}.

\bibitem{chuang1997prescription}
\bibinfo{author}{Chuang, I.~L.} \& \bibinfo{author}{Nielsen, M.~A.}
\newblock \bibinfo{title}{Prescription for experimental determination of the
  dynamics of a quantum black box}.
\newblock \emph{\bibinfo{journal}{Journal of Modern Optics}}
  \textbf{\bibinfo{volume}{44}}, \bibinfo{pages}{2455--2467}
  (\bibinfo{year}{1997}).
\newblock
  \urlprefix\url{https://www.tandfonline.com/doi/abs/10.1080/09500349708231894}.
\newblock
  \eprint{https://www.tandfonline.com/doi/pdf/10.1080/09500349708231894}.

\bibitem{Poyatos97complete}
\bibinfo{author}{Poyatos, J.~F.}, \bibinfo{author}{Cirac, J.~I.} \&
  \bibinfo{author}{Zoller, P.}
\newblock \bibinfo{title}{Complete characterization of a quantum process: The
  two-bit quantum gate}.
\newblock \emph{\bibinfo{journal}{Phys. Rev. Lett.}}
  \textbf{\bibinfo{volume}{78}}, \bibinfo{pages}{390--393}
  (\bibinfo{year}{1997}).
\newblock \urlprefix\url{https://link.aps.org/doi/10.1103/PhysRevLett.78.390}.

\bibitem{d2003quantum}
\bibinfo{author}{D'Ariano, G.~M.}, \bibinfo{author}{Paris, M.~G.} \&
  \bibinfo{author}{Sacchi, M.~F.}
\newblock \bibinfo{title}{Quantum tomography}.
\newblock \emph{\bibinfo{journal}{Advances in Imaging and Electron Physics}}
  \textbf{\bibinfo{volume}{128}}, \bibinfo{pages}{206--309}
  (\bibinfo{year}{2003}).

\bibitem{Mohseni08quantum}
\bibinfo{author}{Mohseni, M.}, \bibinfo{author}{Rezakhani, A.~T.} \&
  \bibinfo{author}{Lidar, D.~A.}
\newblock \bibinfo{title}{Quantum-process tomography: Resource analysis of
  different strategies}.
\newblock \emph{\bibinfo{journal}{Phys. Rev. A}} \textbf{\bibinfo{volume}{77}},
  \bibinfo{pages}{032322} (\bibinfo{year}{2008}).
\newblock \urlprefix\url{https://link.aps.org/doi/10.1103/PhysRevA.77.032322}.

\bibitem{Pogorelov17experimental}
\bibinfo{author}{Pogorelov, I.~A.} \emph{et~al.}
\newblock \bibinfo{title}{Experimental adaptive process tomography}.
\newblock \emph{\bibinfo{journal}{Phys. Rev. A}} \textbf{\bibinfo{volume}{95}},
  \bibinfo{pages}{012302} (\bibinfo{year}{2017}).
\newblock \urlprefix\url{https://link.aps.org/doi/10.1103/PhysRevA.95.012302}.

\bibitem{Granade_2017}
\bibinfo{author}{Granade, C.}, \bibinfo{author}{Ferrie, C.} \&
  \bibinfo{author}{Flammia, S.~T.}
\newblock \bibinfo{title}{Practical adaptive quantum tomography}.
\newblock \emph{\bibinfo{journal}{New Journal of Physics}}
  \textbf{\bibinfo{volume}{19}}, \bibinfo{pages}{113017}
  (\bibinfo{year}{2017}).
\newblock \urlprefix\url{https://doi.org/10.1088%2F1367-2630%2Faa8fe6}.

\bibitem{PhysRevA.90.012110}
\bibinfo{author}{Baldwin, C.~H.}, \bibinfo{author}{Kalev, A.} \&
  \bibinfo{author}{Deutsch, I.~H.}
\newblock \bibinfo{title}{Quantum process tomography of unitary and
  near-unitary maps}.
\newblock \emph{\bibinfo{journal}{Phys. Rev. A}} \textbf{\bibinfo{volume}{90}},
  \bibinfo{pages}{012110} (\bibinfo{year}{2014}).
\newblock \urlprefix\url{https://link.aps.org/doi/10.1103/PhysRevA.90.012110}.

\bibitem{spall1992multivariate}
\bibinfo{author}{Spall, J.~C.} \emph{et~al.}
\newblock \bibinfo{title}{Multivariate stochastic approximation using a
  simultaneous perturbation gradient approximation}.
\newblock \emph{\bibinfo{journal}{IEEE transactions on automatic control}}
  \textbf{\bibinfo{volume}{37}}, \bibinfo{pages}{332--341}
  (\bibinfo{year}{1992}).

\bibitem{simon1990minimal}
\bibinfo{author}{Simon, R.} \& \bibinfo{author}{Mukunda, N.}
\newblock \bibinfo{title}{Minimal three-component su (2) gadget for
  polarization optics}.
\newblock \emph{\bibinfo{journal}{Physics Letters A}}
  \textbf{\bibinfo{volume}{143}}, \bibinfo{pages}{165--169}
  (\bibinfo{year}{1990}).

\bibitem{simon2012hamilton}
\bibinfo{author}{Simon, B.~N.}, \bibinfo{author}{Chandrashekar, C.~M.} \&
  \bibinfo{author}{Simon, S.}
\newblock \bibinfo{title}{Hamilton's turns as a visual tool kit for designing
  single-qubit unitary gates}.
\newblock \emph{\bibinfo{journal}{Phys. Rev. A}} \textbf{\bibinfo{volume}{85}},
  \bibinfo{pages}{022323} (\bibinfo{year}{2012}).
\newblock \urlprefix\url{https://link.aps.org/doi/10.1103/PhysRevA.85.022323}.

\bibitem{Kwia99ultrabright}
\bibinfo{author}{Kwiat, P.~G.}, \bibinfo{author}{Waks, E.},
  \bibinfo{author}{White, A.~G.}, \bibinfo{author}{Appelbaum, I.} \&
  \bibinfo{author}{Eberhard, P.~H.}
\newblock \bibinfo{title}{Ultrabright source of polarization-entangled
  photons}.
\newblock \emph{\bibinfo{journal}{Phys. Rev. A}} \textbf{\bibinfo{volume}{60}},
  \bibinfo{pages}{R773--R776} (\bibinfo{year}{1999}).
\newblock \urlprefix\url{https://link.aps.org/doi/10.1103/PhysRevA.60.R773}.

\end{thebibliography}

\end{document}